\documentclass[useAMS,usenatbib]{mn2e}
\usepackage{amsmath}
\usepackage{amssymb}
\usepackage{graphicx}
\usepackage{times}

\title[Cooling and heating curves]{New composition dependent cooling
  and heating curves for galaxy evolution simulations}

\author[S. De Rijcke et al.]{S. De Rijcke$^{1}$\thanks{E-mail: Sven.Derijcke@Ugent.be},
  J. Schroyen$^{1}$, B. Vandenbroucke$^{1}$, N. Jachowicz$^{2}$,
  J. Decroos$^{1}$, \newauthor A. Cloet-Osselaer$^{1}$,
  M. Koleva$^{1}$\\ $^{1}$Ghent University, Dept. Physics \&
  Astronomy, Krijgslaan 281, S9, B-9000, Ghent, Belgium\\ $^{2}$ Ghent
  University, Dept. Physics \& Astronomy, Proeftuinstraat 86, B-9000,
  Ghent, Belgium}

\begin{document}
\date{}
\pagerange{\pageref{firstpage}--\pageref{lastpage}} \pubyear{2009}

\maketitle

\label{firstpage}

\begin{abstract}
In this paper, we present a new calculation of composition-dependent
radiative cooling and heating curves of low-density gas, intended
primarily for use in numerical simulations of galaxy formation and
evolution. These curves depend on only five parameters:~temperature,
density, redshift, [Fe/H], and [Mg/Fe]. They are easily tabulated and
can be efficiently interpolated during a simulation.

The ionization equilibrium of 14 key elements is determined for
temperatures between $10$~K and $10^9$~K and densities up to
100~amu~cm$^{-3}$ taking into account collisional and radiative
ionization, by the cosmic UV background and an interstellar radiation
field, and by charge-transfer reactions. These elements, ranging from
H to Ni, are the ones most abundantly produced and/or released by
SN{\sc i}a, SN{\sc ii}, and intermediate-mass stars. Self-shielding of
the gas at high densities by neutral Hydrogen is taken into account in
an approximate way by exponentially suppressing the H-ionizing part of
the cosmic UV background for H{\sc i} densities above a threshold
density of $n_{\rm HI, crit}=0.007$~cm$^{-3}$. We discuss how the
ionization equilibrium, and the cooling and heating curves depend on
the physical properties of the gas.

The main advantage of the work presented here is that, within the
confines of a well-defined chemical evolution model and adopting the
ionization equilibrium approximation, it provides accurate cooling and
heating curves for a wide range of physical and chemical gas
properties, including the effects of self-shielding. The latter is key
to resolving the formation of cold, neutral, high-density clouds
suitable for star formation in galaxy simulations.
\end{abstract}

\begin{keywords}
Physical Data and Processes:~atomic processes, hydrodynamics, plasmas,
ISM: general
\end{keywords}

\section{Introduction}\label{intro}

Numerical simulations of galaxy evolution require basic physical input
regarding the (thermo-)dynamical behavior of the interstellar gas. A
crucial ingredient of the energy, or entropy, equation is the cooling
rate of the gas. This quantity is, in principle, a complex function of
the temperature, composition, and irradiation of the gas. An often
used assumption is that the gas is in collisional ionization
equilibrium (CIE). In that case, collisions with free electrons are
deemed solely responsible for keeping atoms ionized. Since both the
recombination rate and the ionization rate are in that case directly
proportional to the electron density, the latter cancels from the
equations and the ionization equilibrium becomes a function of
temperature only (for a given elemental abundance mix). For low gas
densities, each collisional ionization/excitation is followed by a
radiative de-excitation, creating an escaping photon, and the cooling
rate becomes proportional to the density squared (or to Hydrogen
density times electron density, $n_{\rm H} n_{\rm e}$) times a
temperature-dependent function.
\begin{figure*}
\centering
\includegraphics[width=\textwidth]{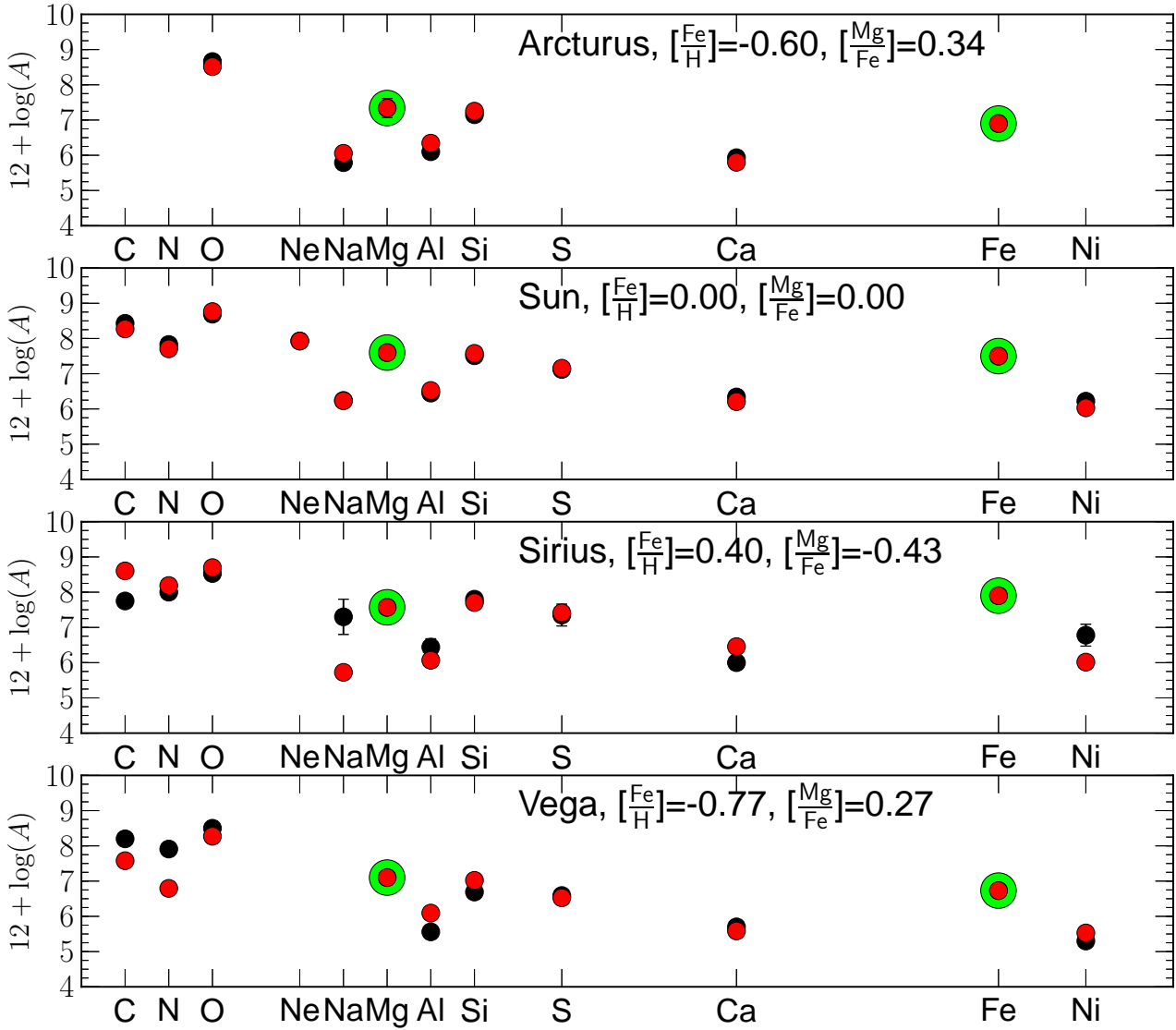}
\caption{The observed metal abundances in the atmospheres of a number
  of well-studied standard stars (black points) and those predicted by
  our chemical evolution model (red points) for the appropriate [Fe/H]
  and [Mg/Fe] values (green points). The abundance, $A$, of an element
  is measured as the ratio of its number density relative to that of
  H.  \label{fig:compsun}}
\end{figure*}

Many state-of-the-art simulation codes
\citep{re09,sa11,sc11,sch11,co12,gd12,ki12} rely on the cooling curves
compiled by \citet{sd93}. The latter authors calculated cooling rates,
excluding a forefactor $n_{\rm H} n_{\rm e}$, as a function of
temperature for a number of metallicities. During a simulation, the
cooling rate of a gas parcel can be rapidly determined by simple
two-dimensional interpolation on these curves. However, while this
work was monumental and has spawned a large volume of literature based
on simulations making use of these curves, one needs to be aware of
the assumptions on which these cooling curves are based and simulators
need to assess whether they can be used for the application at
hand. To be clear:~this is no criticism of the \citet{sd93} cooling
curves.
\begin{itemize}
\item Metallicity is quantified using [Fe/H], the Iron abundance. For
  [Fe/H]$\le -1$ the abundance ratios are taken to reflect those of a
  SN{\sc ii}, with [$\alpha$/Fe]$>0$; for [Fe/H]$=0$ the solar
  abundance ratios are adopted, with [$\alpha$/Fe]$=0$; for other
  metallicities, the abundance ratios are interpolated linearly
  between these two sets of abundance ratios. Hence, {\em adopting the
    \citet{sd93} cooling curves immediately implies adopting the solar
    neighborhood's chemical enrichment history.} In some cases, this
  may not be a good approximation of reality. E.g., dwarf galaxies
  have low metallicities but also low [$\alpha$/Fe] (see
  e.g. \citet{tht09} and references therein). Using the
  low-metallicity \citet{sd93} cooling curves will then quite strongly
  overestimate the cooling contributed by $\alpha$-elements such as O,
  Si, and Mg. The centers of giant elliptical galaxies, on the other
  hand, have high metallicities and high [$\alpha$/Fe]
  \citep{wfg92}. With the \citet{sd93} cooling curves, the
  contribution of the $\alpha$ elements will be strongly
  underestimated.
\item Another issue is whether in the presence of a cosmological UV
  background (UVB) CIE is still an acceptable approximation. The UVB
  tends to keep (part of) the Hydrogen and metals ionized, thus
  lowering the fraction of H{\sc i} and raising $n_{\rm e}$. This
  dramatically influences the shape of the cooling curve, as is well
  known \citep{wi09} and as we will also show below.  \citet{tp11}
    compared the net cooling rates computed assuming only CIE with
    those calculated including photo-ionization. These authors showed
    that the equilibrium temperature of the gas could be off by an
    order of magnitude at low densities and high metallicities when
    using CIE.
\item Simulations nowadays achieve sufficiently high spatial
  resolution to be able to follow the formation of cold, dense
  star-forming clouds. This requires extending the \citet{sd93}
  cooling curves to temperatures below $10^4$~K. An often used
  extension is the set of cooling curves of \citet{ma07}. These
  authors calculate the level populations of Fe{\sc ii}, O{\sc i},
  Si{\sc ii}, and C{\sc ii} and the cooling rates due to the low-lying
  finestructure emission lines of these ions. These level populations
  are set by collisions with free electrons and thus by the ionization
  fraction of the gas. Again, this is a quantity which is very
  sensitive to the presence of a cosmic UVB. Clearly, the often used
  approach of ``gluing'' the \citet{ma07} cooling curves
  ($10$~K$<T<10^4$~K) to the \citet{sd93} cooling curves
  ($10^4$~K$<T<10^9$~K) will produce unreliable results in simulations
  with a cosmic UVB. 
\end{itemize}

 Recent calculations of cooling curves with CLOUDY, such as
\citet{wi09} (used in the OverWhelmingly Large Simulations project
\citep{sch10}) and \citet{sh10} (used in the simulation code GASOLINE
\citep{sh10,br12}), include the cosmic UVB in determining the
ionization balance. Here, the total cooling rate is written
as the sum of approximately independent terms:~the cooling due to H
and He, the cooling due to metals, and inverse Compton
cooling. \citet{wi09} advocate an element-by-element approach,
necessitating the tracing of the abundances of a set of 11 elements
during a simulation in order to calculate the independent contribution
of each to the total cooling rate. In reality, these terms are linked
by the free electron density, by charge-exchange reactions, and by
other reactions between different elements (such as molecule and dust
particle formation). Moreover, many authors still adopt the solar
abundance ratios and scale the metal cooling rate proportional to
metallicity which, as we have argued, can be expected to be a bad
representation of reality for certain types of stellar systems. 
  Likewise, \citet{ssa08}, and \citet{sm11} only provide cooling
  curves for solar abundance ratios of the heavy elements (H and He
  ions are followed explicitly during simulations). In the former
  paper, the influence of the UVB is not taken into account while the
  latter focuses precisely on this issue. \citet{gh12} incorporate the
  radiation field through four well-chosen normalized photoionization
  rates. This, together with a Taylor expansion of the curves up to
quadratic terms in metallicity, yields an approximation to the cooling
and heating curves with a median error of 10~\% but with (although
very rare) errors of up to a factor of six.

In this paper, we try to improve on several aspects of the existing
cooling curve calculations. 

\section{Numerical details} \label{sect:num}

Below, we give a list of the most prominent ingredients of our
calculations:
\begin{itemize}
\item We adopt a chemical enrichment model that is self-consistent in
  the sense that, in an $N$-body/SPH simulation, gas particles can be
  enriched by stellar particles in only two ways:~fast (by SN{\sc ii}
  and massive intermediate-mass stars (IMS), with $M \sim
  8$~M$_\odot$) and slow (by SN{\sc i}a and less massive IMS). Thus,
  the chemical abundance mix of a gas particle depends solely on the
  ratio of the ``slow'' and ``fast'' contributions. The cooling and
  heating rates can then be tabulated for a small number of different
  ratios of ``slow'' to ``fast'' contributions, covering all
  possibilities that can occur in a simulation.

  SN{\sc ii} yields are taken from \citet{no97} and modified according
  to the prescriptions detailed in \citet{fr04}; IMS yields come from
  \citet{ga05}; SN{\sc i}a yields have been adopted from
  \citet{ts95}. For SN{\sc ii}, the fraction of the initial mass of a
  stellar population that is returned in the form of element $X$ to
  the interstellar medium (ISM) is given by
  \begin{equation}
    y_{X, \rm SNII} = \frac{ \int_{m_{\rm lII}}^{m_{\rm uII}} M_X(m) \phi(m) dm }{\int_{m_{\rm l}}^{m_{\rm u}} m \phi(m) dm }.
  \end{equation}
  Here, $m_{\rm lII}=8$~M$_\odot$ and $m_{\rm uII}=70$~M$_\odot$ are
  the lower and upper bounds of the masses of stars that turn into
  SN{\sc ii}; $m_{\rm l}=0.1$~M$_\odot$ and $m_{\rm u}=70$~M$_\odot$
  are the adopted lower and upper bounds of the masses of stars. The
  mass returned in the form of element $X$ by a star with initial mass
  $m$ is denoted by $M_X(m)$. For the initial-mass function, or IMF,
  denoted by $\phi(m)$, we take the parameterization by \citet{cha}.

  For SN{\sc i}a, the fraction of the initial mass of a stellar
  population that is returned in the form of element $X$ to the ISM is
  given by
  \begin{equation}
    y_{X, \rm SNIa} = A_{\rm Ia} M_X \frac{ \int_{m_{\rm lIa}}^{m_{\rm
          lII}} \phi(m) dm }{\int_{m_{\rm l}}^{m_{\rm u}} m
      \phi(m) dm }.
  \end{equation}
  Here, $m_{\rm lIa}=3$~M$_\odot$ is the lower bound of the masses of
  stars that can turn into SN{\sc i}a and $M_X$ is the SN{\sc i}a
  yield of element $X$. The forefactor $A_{\rm Ia}$ was determined
  by demanding that the calculated ratio of the occurrence of SN{\sc
    i}a to that of SN{\sc ii} reproduces that derived for the solar
  neighborhood by \citet{ts95}:
  \begin{equation}
    \frac{N_{\rm Ia}}{N_{\rm II}} = 0.15 = A_{\rm Ia} \frac{ \int_{m_{\rm lIa}}^{m_{\rm
          lII}} \phi(m) dm }{\int_{m_{\rm lII}}^{m_{\rm uII}} \phi(m) dm }.
  \end{equation}
  
  For IMS, the yield is given analogously by
  \begin{equation}
    y_{X, \rm IMS} = \frac{ \int_{m_{\rm lIMS}}^{m_{\rm uIMS}} M_X(m)
      \phi(m) dm }{\int_{m_{\rm l}}^{m_{\rm u}} m \phi(m) dm },
  \end{equation}
  with $m_{\rm lIMS}=0.8$~M$_\odot$ and $m_{\rm
    uIMS}=8$~M$_\odot$. The yields of elements contributed by the most
  massive IMS, such as ${}^{13}$C and N, are added
  to the corresponding SN{\sc ii} yields. Those of elements produced
  by longer-lived stars are added to the corresponding SN{\sc i}a
  yields. This way, there are two contributions to the yield of a
  given element:~a ``fast'' one (encompassing the contributions from
  SN{\sc ii} and massive IMS) and a ``slow'' one
  (encompassing the contributions from SN{\sc i}a and less massive
  IMS). The abundance of each chemical element in
  a gas parcel is then simply the weighted sum of these two
  contributions.

\begin{table}
\caption{Element yields. \label{tab:yields}}
\begin{center}
  \begin{tabular}{|c|c|c|} \hline
    element & slow & fast \\ \cline{1-3}
O & 0.000136 & 0.000937 \\ 
C & 0.000146 & 0.000143 \\ 
Ne & 2.41e-05 & 0.00013 \\ 
Mg & 1.06e-05 & 6.23e-05 \\ 
Si & 1.66e-05 & 4.89e-05 \\ 
Fe & 2.96e-05 & 1.67e-05 \\ 
S & 9.15e-06 & 1.41e-05 \\
N & 5.87e-05 & 9.65e-06 \\ 
Al & 6.46e-08 & 6.51e-06 \\ 
Na & 4.86e-09 & 3.36e-06 \\ 
Ni & 2.99e-07 & 1.67e-06 \\ 
Ca & 9.99e-07 & 1.64e-06 \\ \cline{1-3}
\end{tabular}
\end{center}
\end{table}

  Since this simple chemical evolution model contains two sets of
  yields, the elemental abundance ratios in a given gas parcel can
  also be quantified by just two numbers. Here, we choose [Fe/H] as a
  tracer of the overall metallicity and [Mg/Fe] as a second
  parameter. Mg is an $\alpha$-element released abundantly by SN{\sc
    ii} explosions but produced in only very small amounts by SN{\sc
    i}a explosions so it is a good tracer of the relative weights of
  the ``fast'' and ``slow'' contributions. Moreover, there are now
  quite advanced techniques available to determine the abundances of
  both Fe and Mg in a stellar population from absorption lines in
  optical  spectra. Obviously, Oxygen would also make a good
  discriminator between the ``fast'' and ``slow'' contributions. We
  present the element yields used in this paper, expressed as a
  fraction of the initial mass of a single stellar population, in
  Table \ref{tab:yields}.

  In Fig. \ref{fig:compsun}, we compare the observed abundances of a
  number of well-studied stars, taken from
  \citet{wo09,gr10,fi10,la11}, with those predicted by our chemical
  evolution model for the appropriate [Fe/H] and [Mg/Fe] values. For
  the sun and Arcturus, the largest discrepancies are about
  $0.2$~dex. For Sirius and Vega, there are larger deviations between
  data and model, although the author-to-author scatter on the
  measured abundances of these stars is, admittedly, quite substantive
  (e.g. the Na abundance in the atmosphere of Sirius varies by more
  than one dex between different authors, see
  e.g. \citet{la11}). Taking into account observational uncertainties
  and genuine cosmic scatter, this is reassuring evidence that the
  simple two-yield chemical evolution model employed here works adequately.

\item The ionization balance, electron density, level populations, and
  cooling rates are calculated self-consistently in the presence of a
  cosmic UVB. Here, we adopt the UVB calculated by \citet{fg09}. We
  use the UV spectra available from this author's
  webpage\footnote{https://www.cfa.harvard.edu/$\sim$cgiguere/UVB.html}
    to calculate the ionization and heating rates of all elements.

\item Stars also generate an interstellar radiation field (ISRF),
  capable of ionizing atoms with small ionizing potentials, such as
  C{\sc i}, Si{\sc i}, Mg{\sc i}, Ca{\sc i}, Ca{\sc ii}, Fe{\sc i},
  etc. even at very low gas temperatures. Since the light of newly
  formed stars had to make its way through the H{\sc ii} regions
  surrounding these stars, it no longer has an H-ionizing
  component. We include the parameterized ISRF of \citet{ma83},
  appropriate for the solar neighborhood. The ISRF's main task is to
  keep the elements mentioned above ionized while its precise form has
  shown to be of little consequence. At low gas densities, the UVB is
  the dominant photo-ionizing radiation field while at high densities,
  where star formation becomes important and neutral Hydrogen can
  shield the gas from the UVB, the ISRF gains importance.

\item Through Hydrogen ionizations and cascade recombinations, the
  H-ionizing portion of diffuse UV radiation impacting on a gas cloud
  is converted into lower-frequency radiation. Thus, for sufficiently
  high densities, gas may become self-shielding against H-ionizing UV
  radiation once Hydrogen recombines. Therefore, self-shielding will
  generally be insignificant for temperatures $T \gtrsim 10^4$~K since
  then Hydrogen is collisionally ionized anyway.

This is not a straightforward problem since it in principle requires
solving the radiative transfer equation. However, one can estimate the
critical H density above which self-shielding can be expected to block
most of the ionizing UV radiation. \citet{tu98} put forward $n_{\rm H}
\sim 0.01$~cm$^{-3}$, \citet{at10} estimate that $n_{\rm
  H}=0.007$~cm$^{-3}$, and \citet{ya11} quote $n_{\rm H} =
0.00634$~cm$^{-3}$. With this cut-off density, it is possible to
reproduce the observed mass- and volume-averaged neutral fraction of
the universe at a redshift $z\sim 6$ \citep{at10}, the H{\sc i} column
density distribution of damped Ly$\alpha$ systems $z = 3$
\citep{na10}, and the Ly$\alpha$ luminosity of forming galaxies
\citep{f10}.

  We have implemented an approximate scheme for self-shielding by
  exponentially suppressing the UV radiation field with frequencies
  above $h\nu=\chi_{\rm HI}$, with $\chi_{\rm HI}$ the Hydrogen
  ionization potential, as
\begin{eqnarray}
  J_\nu(\nu, n_{\rm HI}) &=& J_\nu(\nu)\exp( -n_{\rm HI}/n_{\rm HI, crit} ) \hspace*{1em} h\nu>1\textrm{Ry} \nonumber \\
&=& J_\nu(\nu) \hspace*{9.2em} h\nu\le 1\textrm{Ry},
\end{eqnarray}
with $n_{\rm HI, crit}=0.007$~cm$^{-3}$ and $J_\nu(\nu)$ the original
UV spectrum, as in \citet{at10}. Note that we use the {\em neutral} H
density here, not the {\em total} H density, since it is only the
neutral fraction which is responsible for absorbing H-ionizing UV
radiation.

One could worry that, when the gas at high densities and low
temperatures becomes self-shielding against the external UVB, the
cooling radiation itself may become trapped and be re-absorbed,
affecting the cooling rate and the ionization equilibrium
\citep{gs07}. However, in a self-shielding H{\sc i} cloud below $T
\sim 10^4$~K, only low-energy UV photons unable to photo-heat the
Hydrogen gas are emitted. Moreover, with most of the star-formation
prescriptions currently popular in galaxy evolution and cosmological
simulations \citep{go10,sch11,co12}, such clouds will begin to form
stars before reaching densities exceeding $100$~amu~cm$^{-3}$ and
stellar and supernova feedback will rapidly overwhelm any internal
diffuse radiation field. Therefore, we expect this to be a minor
issue.

\item Charge-exchange reactions can efficiently transfer electrons
  between ions with similar ionization potentials. Given their high
  abundances, H{\sc i} and H{\sc ii} are the ions' most likely
  reaction partners. Some of the ions that play an important role in
  gas cooling below $10^4$~K via fine-structure line emission, such as
  O{\sc i}, are particularly affected by these reactions. We adopt the
  charge-transfer reaction rates for C{\sc i}, C{\sc ii}, O{\sc i-v},
  Si{\sc i-v}, and Fe{\sc i-v} from \citet{kf96,st98,st99} and from
  the online ORNL/UGA Charge Transfer Database for Astrophysics
  (http://www-cfadc.phy.ornl.gov/astro/ps/data/).
\item Charge-exchange reactions are but one example of reactions that
  involve ions of different elements. Other examples are molecule and
  dust particle formation. We opted not to include molecular
    processes in the present work for the following reasons. Judging
    from e.g. \citet{ma07} and \citet{va13}, cooling below $T \sim
    10^3$~K by molecules is dominated by cooling by metals once the
    latter are present at levels $Z \gtrsim 10^{-3} Z_\odot$. However,
    at the low metallicities where H$_2$ cooling might be important,
    the low dust content strongly inhibits the formation of H$_2$ and
    the timescale for the conversion of H{\sc i} to H$_2$ becomes much
    larger than the local free-fall time \citep{kr12}. Hence, in this
    metallicity range, star formation will precede the formation of
    H$_2$. Also, the rates at which dust forms and at which H$_2$
    molecules form via grain catalysis contain many extra free
    parameters (such as the timescales for dust formation, growth, and
    destruction, the dust grain size, the probability for an H atom to
    stick to a grain, the probability that two H atoms on a grain join
    and detach themselves from the grain as a single H$_2$ molecule,
    etc.) that need to be constrained by experiments and observations
    \citep{b13}. Still, by neglecting molecular cooling we might be
    under-estimating the cooling rate at temperatures below $T \sim
    10^3$~K for metallicities below $Z \sim 10^{-3} Z_\odot$.
  However, by not using the ``independent element'' approximation, as
  in \citet{wi09} and \citet{sh10}, to calculate the ionization
  balance and the cooling curves, our approach can straightforwardly
  be extended in future work.
\item Using all these ingredients, we calculate cooling and heating
  curves for the temperature range $10~{\rm K} < T < 10^9$~K. This
  way, the cooling and heating rates are calculated in perfect
  consistence with the ionization equilibrium over a very wide range
  of temperatures. Thus, there is no need to stitch together cooling
  curves from different authors with potentially very different (and
  inconsistent) input physics.
\end{itemize}


For this work, we have extended the capabilities of ChiantiPy, a
Python interface to the CHIANTI atomic database \citep{de09}, available from
{\tt http://chiantipy.sourceforge.net/}. For all ions, we use the
recombination rates, collisional ionization rates, and energy level
populations provided by standard ChiantiPy. Photo-ionization
cross-sections $\sigma_i(\nu)$ are adopted from \citet{ve96} and
integrated over the stellar and cosmic UV backgrounds in order to
obtain the photo-ionization rate
\begin{equation}
\Gamma_i = 4\pi \int_{\nu_i}^\infty \sigma_i(\nu) J_\nu(\nu) \frac{d\nu}{h\nu},
\end{equation}
and the photo-heating rate
\begin{equation}
\dot{q}_i = 4 \pi \int_{\nu_i}^\infty \sigma_i(\nu) J_\nu(\nu)
(h\nu-h\nu_i) \frac{d\nu}{h\nu},
\end{equation}
with $\nu_i$ the ion's ionization threshold. The integral over the
radiation backgrounds is split in two parts:~the part for photon
energies above 1~Ry, which can be suppressed by H{\sc i}
self-shielding, and the part for photon energies below 1~Ry, which is
assumed to be unaffected by self-shielding. For instance, the
photo-ionization rate of an element can then be written as
\begin{eqnarray}
\Gamma_i &=& {\rm e}^{-n_{\rm HI}/n_{\rm HI, crit}}4\pi \int_{1~\rm Ry}^\infty
\sigma_i(\nu) J_\nu(\nu) \frac{d\nu}{h\nu} \nonumber \\
&& \hspace*{2.5em}+ 4\pi \int_{\nu_i}^{1~\rm Ry}
\sigma_i(\nu) J_\nu(\nu) \frac{d\nu}{h\nu}.
\end{eqnarray}

\begin{figure}
\centering
\includegraphics[width=0.485\textwidth]{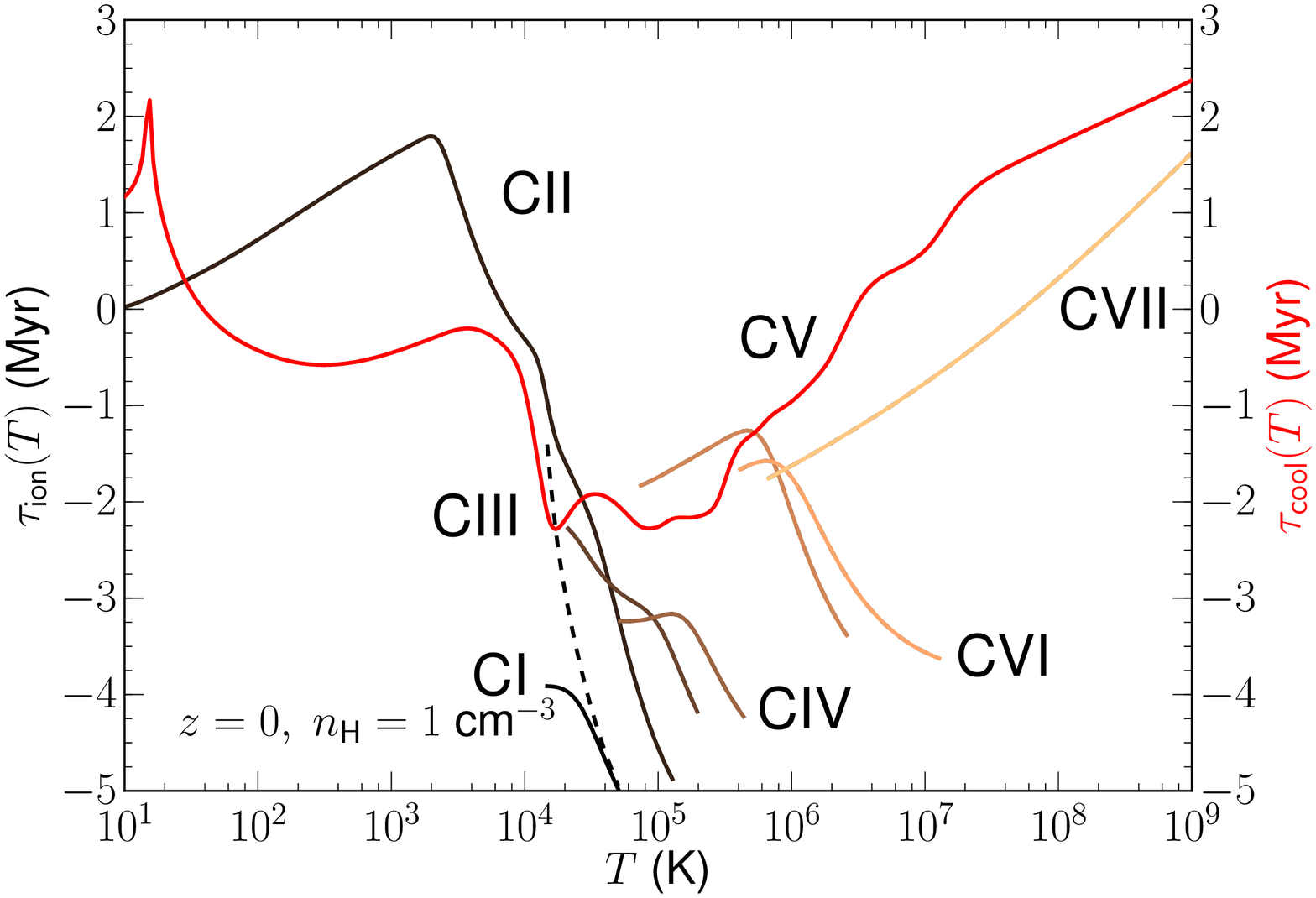}
\includegraphics[width=0.485\textwidth]{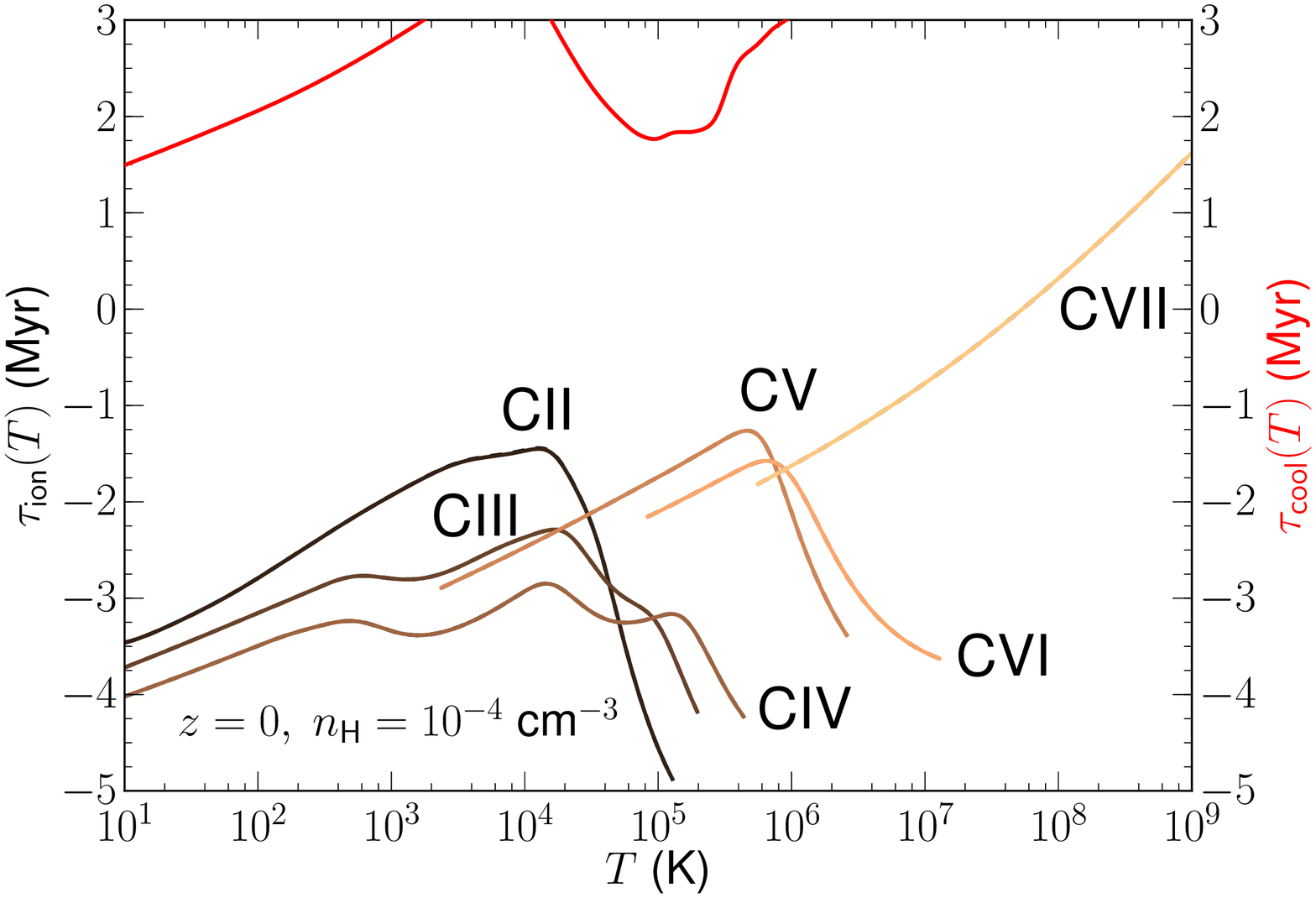}
\includegraphics[width=0.485\textwidth]{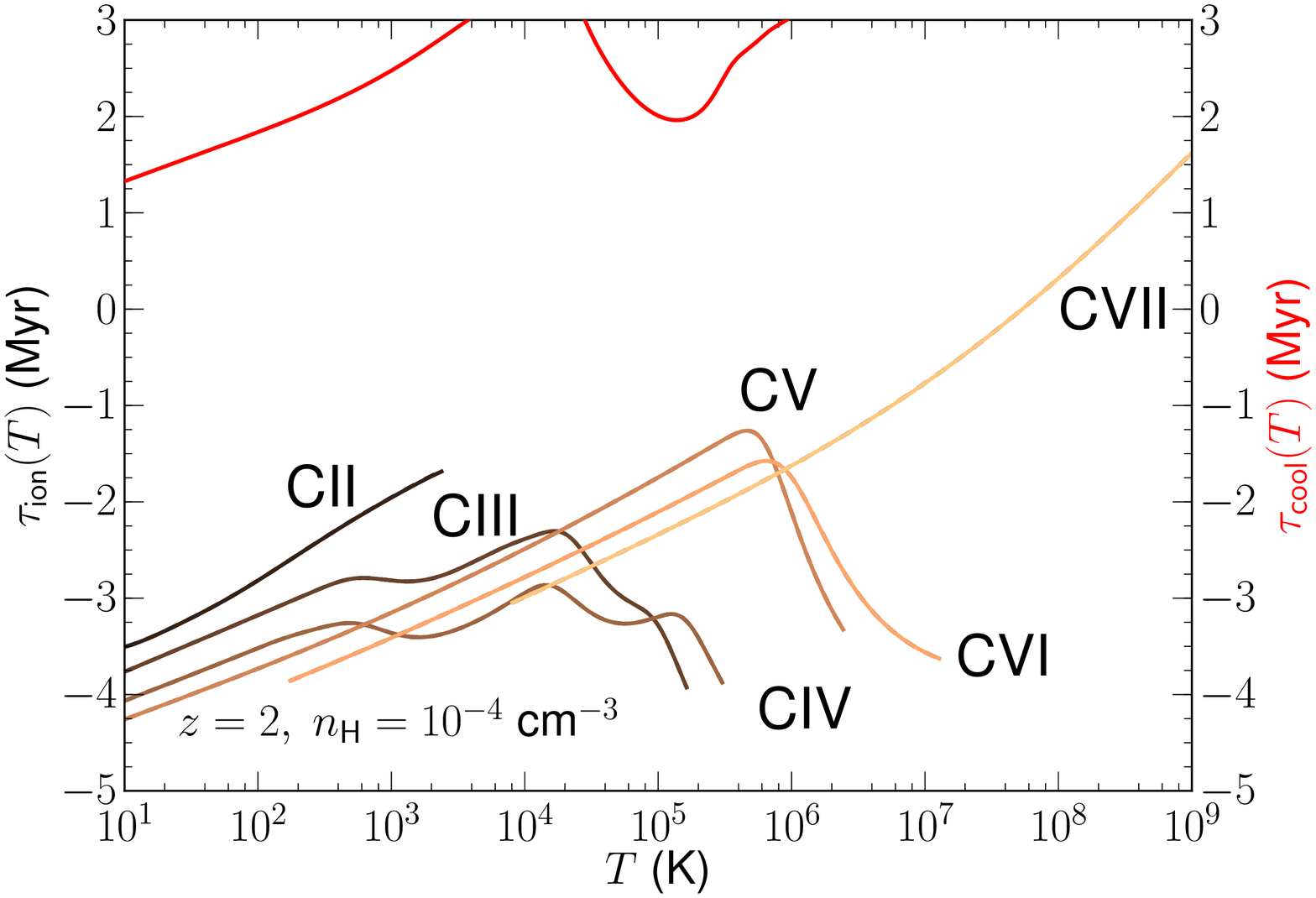}
\caption{Carbon ions ionization timescale, $\tau_{\sf ion}$ (colored
  curves), and cooling timescale, $\tau_{\sf cool}$ (red curve), as a
  function of temperature for different redshifts and densities (as
  indicated in the panels). $\tau_{\sf ion}$ is only plotted if the
  fraction of the corresponding C ion is above $0.001$. The dashed
  curves trace the ionization timescale for pure collisional ionzation
  (this only makes a difference for C{\sc i} at high densities).
\label{fig:timescales}}
\end{figure}

\begin{figure*}
\includegraphics[width=1.02\textwidth]{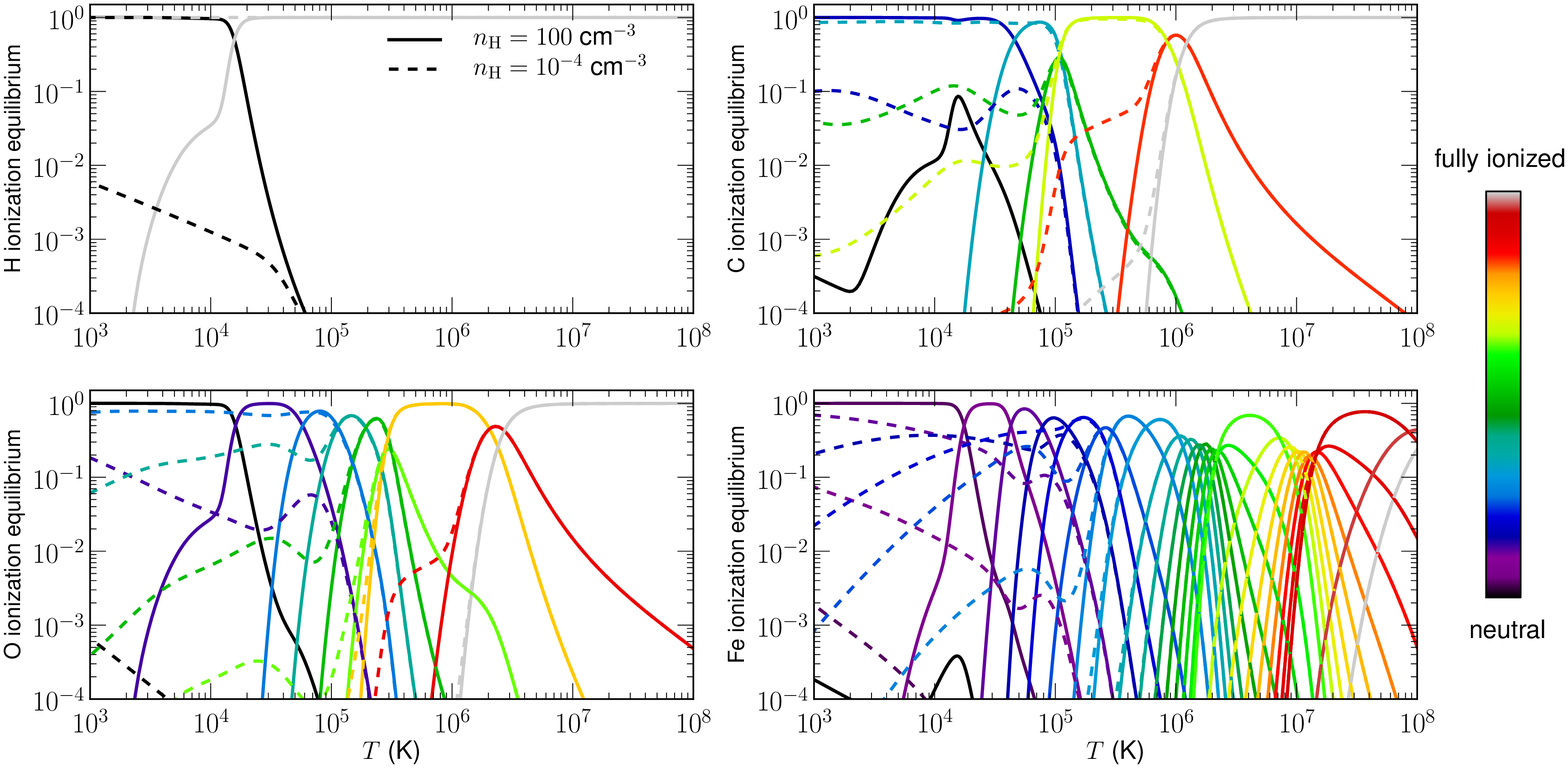}
\caption{Ionization equilibrium of Hydrogen (H), Oxygen (O), Carbon
  (C), and Iron (Fe) as a function of temperature, calculated for the
  UVB at redshift $z=0$ and a density of $n_{\sf H}=100$~cm$^{-3}$ (full
  lines) and $n_{\sf H}=10^{-4}$~cm$^{-3}$ (dashed lines). The color scale
  indicates the ionization stage of the various ions.
\label{fig:ioneq}}
\end{figure*}
\begin{figure*}
\includegraphics[width=1.02\textwidth]{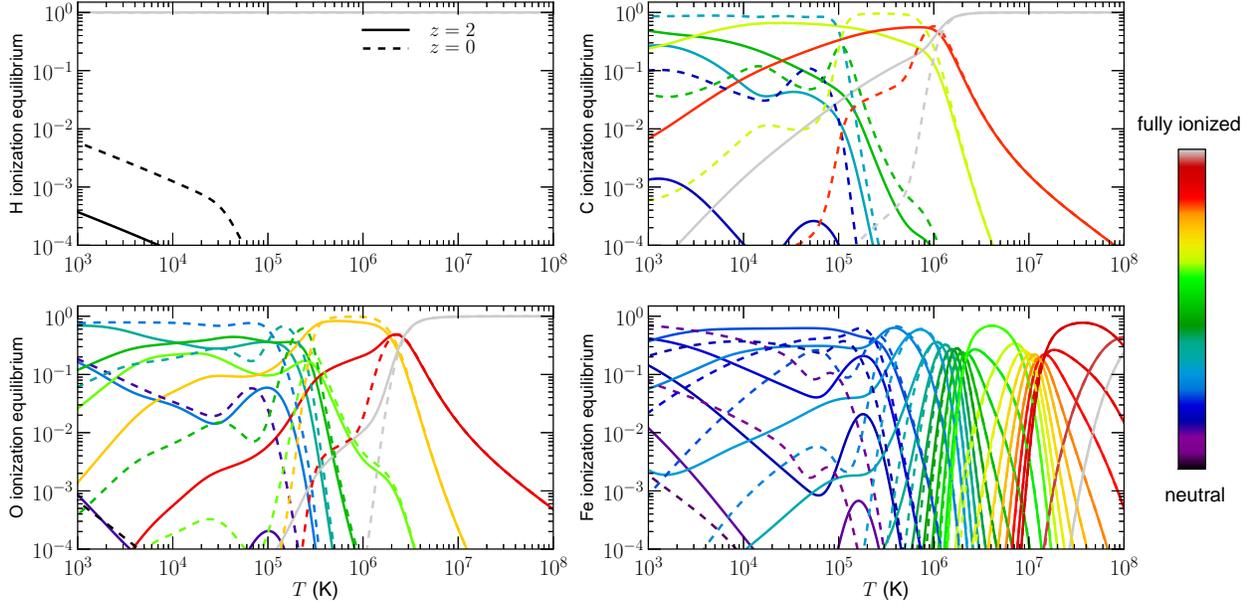}
\caption{Ionization equilibrium of Hydrogen (H), Oxygen (O), Carbon
  (C), and Iron (Fe) as a function of temperature, calculated for a
  density of $n_{\sf H}=10^{-4}$~cm$^{-3}$ for a $z=2$ UVB (full lines) and
  $z=0$ UVB (dashed lines). The color scale indicates the ionization
  stage of the various ions.
\label{fig:ioneqz}}
\end{figure*}

For a given temperature, the ionization equilibrium, i.e. the density
of the $r$-times ionized ion of any element with atomic number $Z$,
denoted by $n_{Z,r}$, is found employing a multi-dimensional
Newton-Raphson technique using back-tracking. At each iteration, the
electron density is given by
\begin{equation}
n_e = \sum_{Z\ge 1} \sum_{r= 0}^Z r n_{Z,r}.
\end{equation}
The general equilibrium condition then becomes
\begin{eqnarray}
\lefteqn{ n_{Z,r} \biggl[ n_e {\sf Rec}_{Z,r}(T) + \sum_c n_c {\sf
      Ion}^{c}_{Z,r}(T) + \Gamma_{Z,r}}  \nonumber \\&& \hspace*{4em} +
    n_{\rm HI}{\sf CT}^{\rm HI}_{Z,r}(T)+ n_{\rm HII}{\sf CT}^{\rm
      HII}_{Z,r}(T) \biggr] \nonumber \\&&= n_{Z,r+1} \left[ n_e {\sf
      Rec}_{Z,r+1}(T) + n_{\rm HI} {\sf CT}^{\rm HI}_{Z,r+1}(T) \right]
  \nonumber \\&&+ n_{Z,r-1} \biggl[ \sum_c n_c {\sf Ion}^{c}_{Z,r-1}(T)
    + n_{\rm HII} {\sf CT}^{\rm
    HII}_{Z,r-1}(T) \nonumber \\&&   \hspace*{4em} + \Gamma_{Z,r-1} \biggr].
\end{eqnarray}

Here, ${\sf Rec}$ indicates the ionic recombination rate, ${\sf
  Ion}^c$ represents the collisional ionization rate with collisional
partner $c$ (which could be electrons, protons, Hydrogen atoms,
\ldots), and ${\sf CT}$ stand for charge-transfer reaction rates (some
of which are obviously zero, such as the reaction rate between H{\sc
  i} and a neutral atom). Given the strong, in this case exponential,
dependence of the self-shielding on the neutral Hydrogen fraction this
is clearly a very non-linear set of equations. 

Moreover, the UVB, at least at sufficiently low gas densities, can be
expected to keep a large fraction of the Hydrogen gas ionized, thereby
suppressing the bound-bound and free-bound cooling contributed by
Hydrogen. As a consequence, the pronounced peak in the CIE cooling
rate around $T \sim 10^4$~K which is caused by Hydrogen, will be
absent. Since non-equilibrium cooling generally leads to
over-ionization compared with CIE, it also tends to suppress the H
cooling peak. Hence, the ionization equilibrium assumption will have a
smaller impact on cooling rates calculated in the presence of an UVB
than on cooling rates calculated assuming CIE \citep{wi09}. As
  check on the assumption of ionization equilibrium, we calculated the
  ionization timescales, $\tau_{\sf ion}$, of e.g. the Carbon ions to
  be compared with the cooling timescale, $\tau_{\sf cool}$. The
  former is given by
\begin{eqnarray}
\frac{1}{\tau_{\sf ion}} &\approx& n_e {\sf Rec}_{Z,r}(T) + \sum_c n_c {\sf
      Ion}^{c}_{Z,r}(T) + \Gamma_{Z,r}  \nonumber \\ && \hspace*{4em} +
    n_{\rm HI}{\sf CT}^{\rm HI}_{Z,r}(T)+ n_{\rm HII}{\sf CT}^{\rm
      HII}_{Z,r}(T)
\end{eqnarray}
while for the latter we use
\begin{equation}
\tau_{\sf cool} = \frac{3}{2} \frac{nkT}{\left| \Lambda(T)-{\cal H}(T) \right| }
= \frac{3}{2} \frac{nkT}{ \Lambda_{\sf net}(T)  }
\end{equation}
with $\Lambda(T)$ the cooling rate and ${\cal H}(T)$ the heating rate.
The ionization equilibrium approximation is valid if $\tau_{\sf cool}
\gg \tau_{\sf ion}$ for all ions. Judging from
Fig. \ref{fig:timescales}, this constraint is more easily fulfilled at
low densities, when the UVB irradiates the gas unimpeded and keeps
most of the Hydrogen ionized. Hence, the net cooling rate
$\Lambda_{\sf net}(T)$ is small. At higher densities the UVB is
attenuated, Hydrogen recombines and $\Lambda_{\sf net}(T)$ is
large. This makes it much harder for this constraint to be fulfilled,
especially at lower temperatures. This is a caveat that should, of
course, be kept in mind when using any set of cooling and heating
tables calculated assuming ionization equilibrium. 

\section{Ionization balance}

\subsection{Density dependence}

In Fig. \ref{fig:ioneq}, we show the ionization equilibrium (i.e. the
fraction of the atoms of a given element that come in the form of a
given ion) of Hydrogen (H), Oxygen (O), Carbon (C), and Iron (Fe) as a
function of temperature, calculated for the UVB at redshift $z=0$ and
a density of $n_{\sf H}=100$~cm$^{-3}$ (full lines) and $n_{\sf
  H}=10^{-4}$~cm$^{-3}$ (dashed lines).  At densities above the
self-shielding density threshold, Hydrogen can recombine at
temperatures below about 20,000~K, thus shielding the gas from the
photo-ionizing UVB. Likewise, Oxygen, with its first ionization
potential very close to that of Hydrogen, recombines to its neutral
form. Only elements with first ionization potentials smaller than that
of Hydrogen, such as Carbon and Iron in this example, remain once
ionized below $\sim 10,000$~K.

At low densities, below the self-shielding density threshold of
Hydrogen, the ionizing UVB can flood the gas unimpeded, keeping over
99~\% of the Hydrogen ionized. This completely erases the contribution
of Hydrogen to the cooling via its free-bound and bound-bound
transitions. In this example the UVB affects essentially all ions of
Carbon and Oxygen, with C{\sc iii}, O{\sc iii}, and Fe{\sc iv} the
most common ionization stages of these elements at low
temperatures. The high abundance of free photo-electrons provides
extra cooling through radiative free-bound transitions, more than
making up for the lack of efficient coolants such as the C{\sc ii} and
Fe{\sc ii} infrared fine-structure lines at low temperatures (see
below, in section \ref{sect:cooling}).

\begin{figure}
\centering
\includegraphics[width=0.49\textwidth]{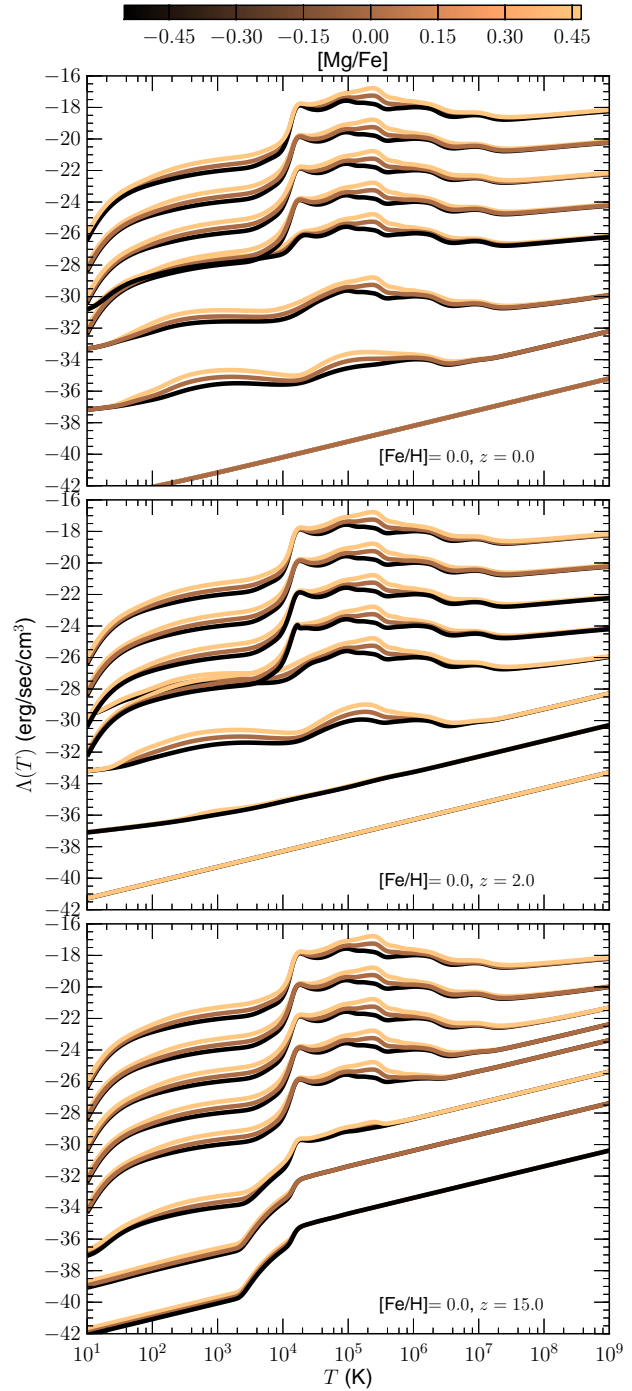}
\caption{Logarithm of the cooling rate as a function of temperature
  for [Fe/H]$=0.0$, and for the $z=0$ (top panel), $z=2$ (middle
  panel), and $z=15$ (bottom panel) cosmic UVB, plotted for different
  densities (from the top curve downwards:~$n_{\sf H}=100$, $10$, $1$,
  $10^{-1}$, $10^{-2}$, $10^{-4}$, $10^{-6}$, $10^{-9}$~cm$^{-3}$)
  and [Mg/Fe]-values (color code).
\label{fig:cooling_dens_z2_feh0.eps}}
\end{figure}

\subsection{Redshift dependence}

For densities below the Hydrogen self-shielding density threshold, the
ionizing strength of the UVB, which varies significantly with
redshift, has a profound impact on the ionization balance. In
Fig. \ref{fig:ioneqz}, we show the ionization equilibrium of Hydrogen
(H), Oxygen (O), Carbon (C), and Iron (Fe) as a function of
temperature, calculated for a gas with a density of $n_{\sf
  H}=10^{-4}$~cm$^{-3}$ and subjected to a $z=2$ UVB (full lines) and
a $z=0$ UVB (dashed lines). Clearly, the stronger $z=2$ UVB leads to
more ionization of the various elements with C{\sc iii} and C{\sc iv}
the dominant Carbon ions at low temperatures while Oxygen is found
predominatly in its O{\sc iv} and O{\sc v} forms. Below $10^5$~K Iron
exists mostly as Fe{\sc vii} and Fe{\sc viii}.

\section{Cooling} \label{sect:cooling}

\subsection{Density dependence}

In the top panel of Fig. \ref{fig:cooling_dens_z2_feh0.eps}, we
compare cooling curves, denoted by $\Lambda(T)$, calculated using the
$z=0$ UVB, [Fe/H]=0, and for different densities between $n_{\rm
  H}=10^{-9}$~cm$^{-3}$ and $n_{\rm H}=100$~cm$^{-3}$, as indicated in
the figure.

At the lowest densities, inverse Compton scattering dominates the
cooling rate at all temperatures and the cooling curve is virtually
featureless. At higher densities, in the range $10^{-9} < n_{\rm H}
\ll 10^{-2}$~cm$^{-3}$, cooling via radiative free-bound and
bound-bound transitions becomes important at temperatures below $\sim
10^7$~K. At these densities, Hydrogen is essentially completely
photo-ionized by the UVB and does not contribute to the cooling rate
via these transitions. The many free electrons keep the rate of
free-bound transitions high even at the lowest temperatures considered
here. At the highest temperatures, upwards of $\sim 10^7$~K, the
cooling curve's temperature slowly changes from the $\Lambda \propto
T$ behaviour of inverse Compton scattering to the less steep $\Lambda
\propto \sqrt{T}$ behaviour that is characteristic of free-free
transitions.  At the highest densities, for $n_{\rm H} \gtrsim
10^{-2}$~cm$^{-3}$, Hydrogen is able to recombine and shield the other
elements from the UVB. Consequently, the $10,000-20,000$~K peak in the
cooling rate contributed by free-bound and bound-bound transitions of
Hydrogen appears. As a result of the strong reduction of the free
electron density, the cooling rate is likewise strongly reduced below
$T\sim 100$~K. Below $10^4$~K, cooling via infrared finestructure
emission lines, predominantly the 157.7~$\mu$m line of C{\sc ii}
dominates.

\subsection{Redshift dependence}
\begin{figure*}
\centering
\includegraphics[width=0.49\textwidth]{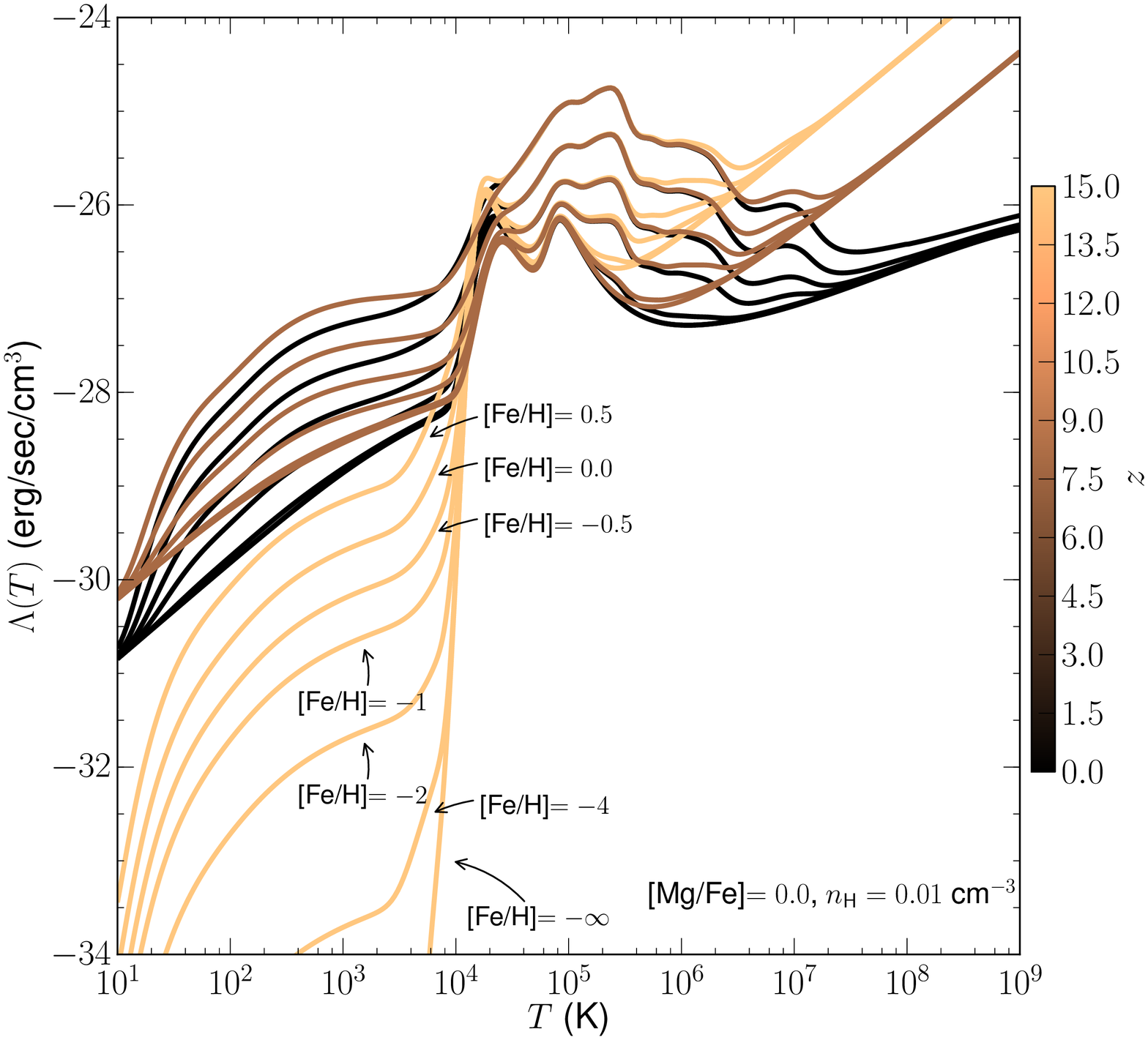}
\includegraphics[width=0.49\textwidth]{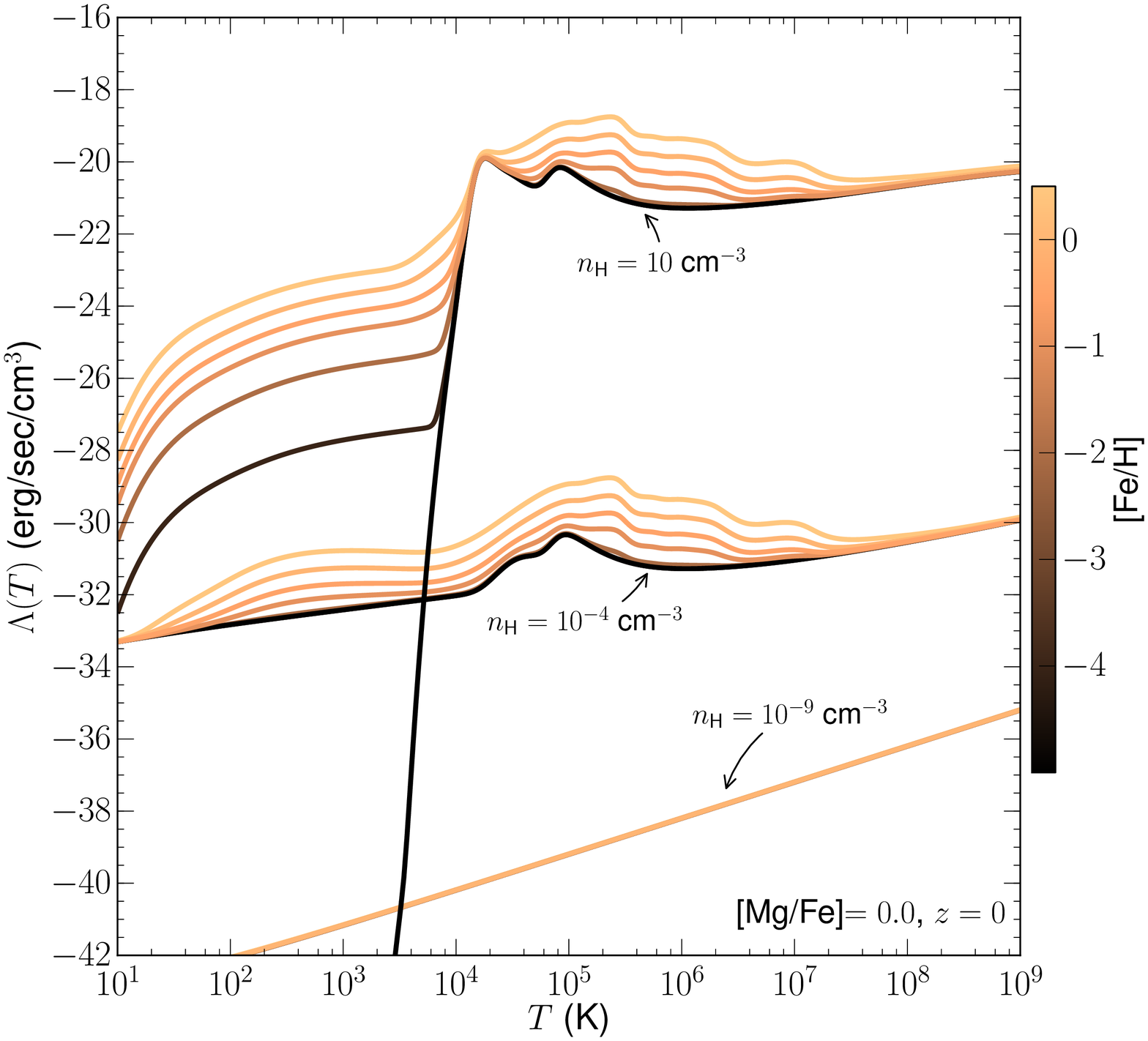}
\caption{Left panel:~logarithm of the cooling rate as a function of
  temperature for Mg-abundance [Mg/Fe]$=0.0$, density $n_{\sf
    H}=0.01$~cm$^{-3}$, for redshifts $z=0$, $z=8$, and $z=15$ (color
  code), and for metallicities between [Fe/H]$=-\infty$ and
  [Fe/H]$=0.5$ (as indicated in the figure). \newline Right panel:~the
  logarithm of the cooling rate as a function of temperature for
  Mg-abundance [Mg/Fe]$=0.0$, redshift $z=0$, densities $n_{\sf
    H}=10^{-9}$~cm$^{-3}$, $n_{\sf H}=10^{-4}$~cm$^{-3}$, and $n_{\sf
    H}=10$~cm$^{-3}$, and for different metallicities (color
  code). The bottom, black curve of each series is calculated for
  [Fe/H]$=-\infty$.
\label{fig:cooling_dens_z0.0_feh0.eps}}
\end{figure*}



The cosmological redshift enters the picture in two distinct ways:~on
the one hand, it controls the strength of the UVB, and on the other,
it determines the contribution to the cooling rate from inverse
Compton scattering through its $(1+z)^4$-dependence. For the
zero-strength $z=15$ UVB, H always fully recombines below $T \sim
10^4$~K and the cooling rate plummets by 4 orders of magnitude, as is
evident in the bottom panel of
Fig. \ref{fig:cooling_dens_z2_feh0.eps}. The only contribution to the
inverse Compton scattering cooling rate now comes from the handful of
ionization electrons from ions with small ionizing potentials, such as
C{\sc i}, Fe{\sc i}, etc.

The stronger the UVB, the more highly ionized the different elements
are. This lack of lowly ionized species leads to a dramatic decrease of
the cooling rate due to free-bound and bound-bound emission (compare
e.g. the cooling curves for densities $n_{\sf H}=10^{-6\rightarrow
  -4}$~cm$^{-3}$ between $z=0$ and $z=2$). 

\subsection{[Fe/H] dependence}

In Fig. \ref{fig:cooling_dens_z0.0_feh0.eps}, we compare the cooling
rates at metallicities between [Fe/H]$=-\infty$ and [Fe/H]$=0.5$ (as
indicated in the figure), calculated for Mg-abundance [Mg/Fe]$=0.0$,
density $n_{\sf H}=0.01$~cm$^{-3}$, and for redshifts $z=0$, $z=8$,
and $z=15$. 

At high temperatures, inverse Compton scattering and free-free
interactions dominate the cooling rate at high redshift. In this
regime, the metallicity only affects the density of free electrons
and, as a result, the cooling curves are not very sensitive to
metallicity. At temperatures $T \lesssim 10^7$~K, partially ionized
atoms can exist and free-bound and bound-bound transitions contribute
greatly to the cooling rate. Likewise, for $T \lesssim 10^4$~K, the
cooling rate via infrared fine-structure lines is a strong function of
metallicity since in this regime the cooling depends crucially on the
presence of a few key ions. This is especially true when the UVB is
very weak or even absent (as at $z=15$). In that case, Hydrogen fully
recombines below $T \sim 10^4$~K, consuming all free electrons, and
the cooling rate drops sharply. The cooling contributed by infrared
fine-structure lines of metal ions can then make a huge difference.

\subsection{[Mg/Fe] dependence}

In Fig. \ref{fig:cooling_dens_z2_feh0.eps}, the cooling curves are
color coded according to their [Mg/Fe]-value. The amount of
$\alpha$-enhancement clearly has a great impact on the cooling rate in
those temperature ranges where key ions of the $\alpha$-elements
contribute free-bound and bound-bound cooling. Around $T \sim
200,000$~K, depending on the abundance of O and Ne, the cooling rate
can vary by up to an order or magnitude. Around $T \sim 10^6$~K, Si
lets its presence be felt. In the range $10^2 \lesssim T \lesssim
10^4$~K, infrared fine-structure emission lines from $\alpha$-element
ions such as O{\sc i} and Si{\sc ii} contribute to the cooling rate
\citep{ma07} and can make an order of magnitude difference depending
on whether a gas parcel has been enriched only the by SN{\sc i}a and
low-mass stars (the ``slow'' contribution to the yield) or only by
SN{\sc ii} and massive stars (the ``fast'' contribution to the
yield). Since the ``fast'' and ``slow'' yields of Carbon happen to be
very similar, the cooling rate is relatively unsensitive to [Mg/Fe]
below $100$~K. For a very weak or absent UVB, ions of
$\alpha$-elements with small first ionization potentials, ionized by
the ISRF, contribute free electrons and slightly raise the cooling
rate below $T \sim 10^4$~K (this is most noticeable in the bottom
panel of Fig. \ref{fig:cooling_dens_z2_feh0.eps}).

\section{Heating}

The heating rate is a very strong function of gas density and
metallicity. Both parameters determine the density of partially
ionized atoms capable of absorbing energy from the UVB through further
ionization while the former, moreover, sets the amount of
self-shielding. At low densities, the heating rate is essentially a
 monotonically declining function of temperature:~the higher the
temperature the lower the densities of the lowly-ionized species that
absorb heat most efficiently. The heating rate also increases with
increasing metallicity since this obviously raises the number of heat
absorbing ions.

\begin{figure*}
\includegraphics[width=\textwidth]{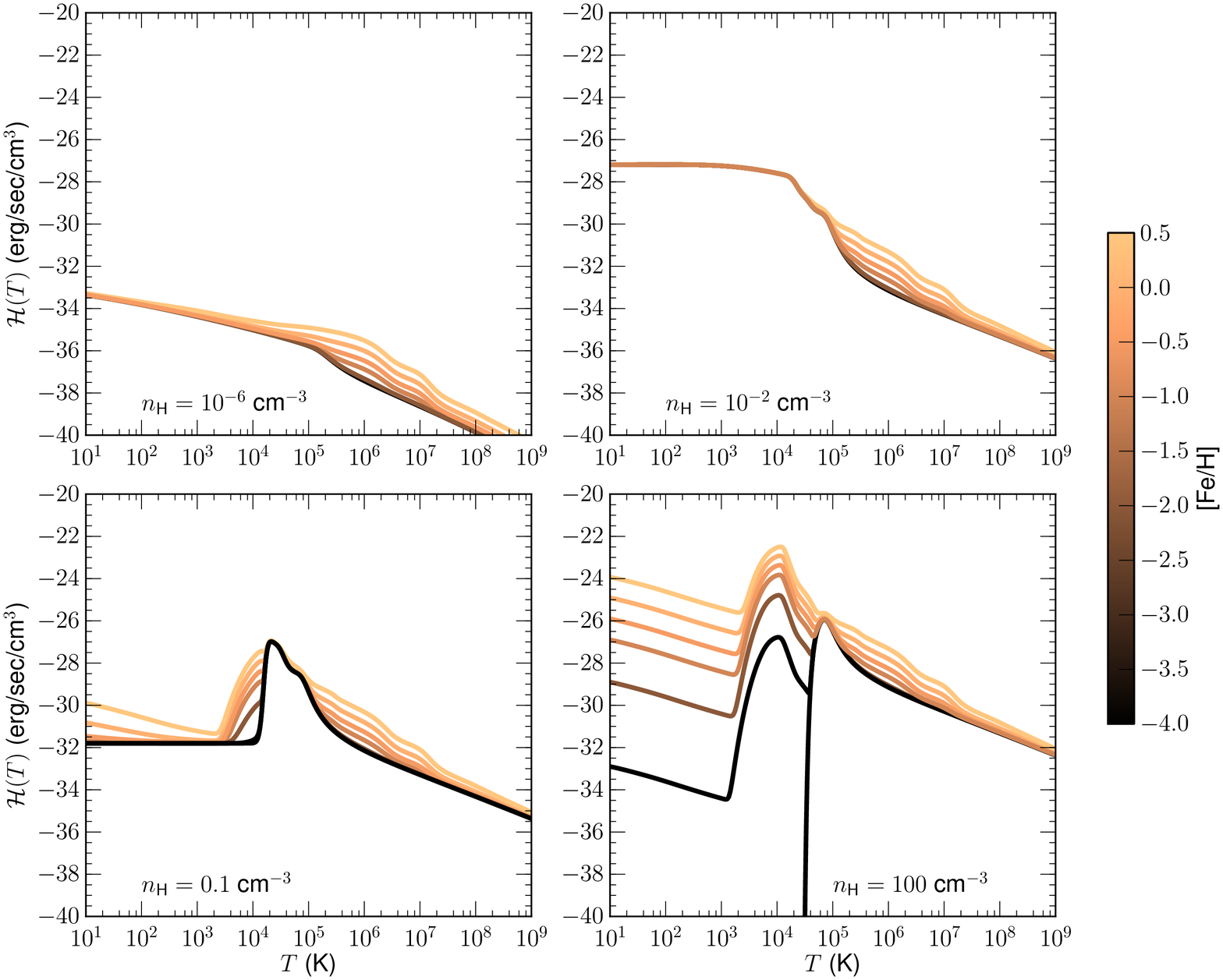}
\caption{Logarithm of the heating rate as a function of temperature
  for Mg-abundance [Mg/Fe]$=0.0$, redshift $z=0$, Fe-abundance [Fe/H]
  (color code), and different gas densities (indicated in the panels).
\label{fig:heating}}
\end{figure*}

At high densities, the heating rate shows a much more complex
behavior. While the densities of near-neutral species increase towards
lower temperatures, potentially raising the heat-absorbing
capabilities of the gas, the self-shielding by neutral Hydrogen
suppresses the UVB. This can lead to a plateau in the heating rate for
temperatures below $10^4$~K, see e.g. at $n_{\sf H}=0.1$~cm$^{-3}$ in
Fig. \ref{fig:heating}. This plateau is simply the heating rate of a
fully neutral Hydrogen gas irradiated by a strongly reduced UVB. For
higher metallicities, lowly ionized metals provide a small amount of
extra heating above this plateau.

 At the highest densities, the Hydrogen-ionizing part of the UVB is
almost completely suppressed and other sources of heating, such as the
ISRF, become significant. Since Hydrogen cannot extract heat from the
ISRF employed in our calculations, the heating rate drops to zero for
[Fe/H]$=-\infty$ when Hydrogen recombines. For non-zero metallicities,
the ISRF can ionize and thus heat those elements that have ionization
potentials smaller than that of Hydrogen, explaining the complex
behavior of the heating rate between $10^3$ and $10^4$~K.

\begin{figure}
\includegraphics[width=0.5\textwidth]{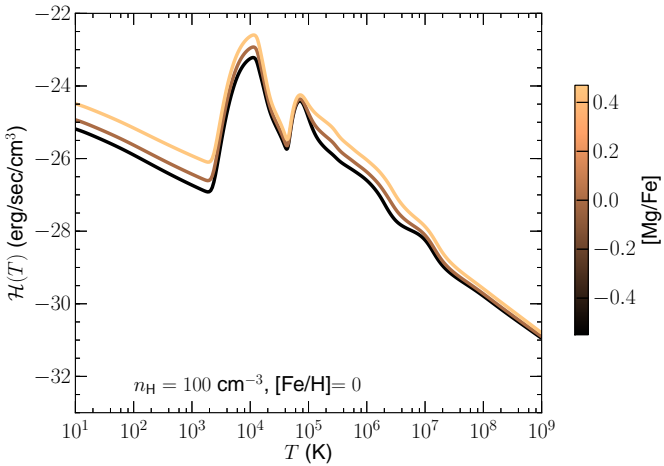}
\caption{Logarithm of the heating rate as a function of temperature
  for Fe-abundance [Fe/H]$=0.0$, redshift $z=0$, gas density $n_{\sf
    H}=100$~cm$^{-3}$, and different [Mg/Fe] ratios (color code).
\label{fig:heatingmgfe}}
\end{figure}

The [Mg/Fe]-dependence of the heating rates is much weaker than their
metallicity dependence. As an example, we show the heating rate as a
function of temperature for Fe-abundance [Fe/H]$=0.0$, redshift $z=0$,
gas density $n_{\sf H}=100$~cm$^{-3}$, and different [Mg/Fe] ratios in
Fig. \ref{fig:heatingmgfe}. While [Mg/Fe] varies with one dex, the
heating rate changes by a factor of 5 at most.

\begin{figure}
\centering
\includegraphics[width=0.5\textwidth]{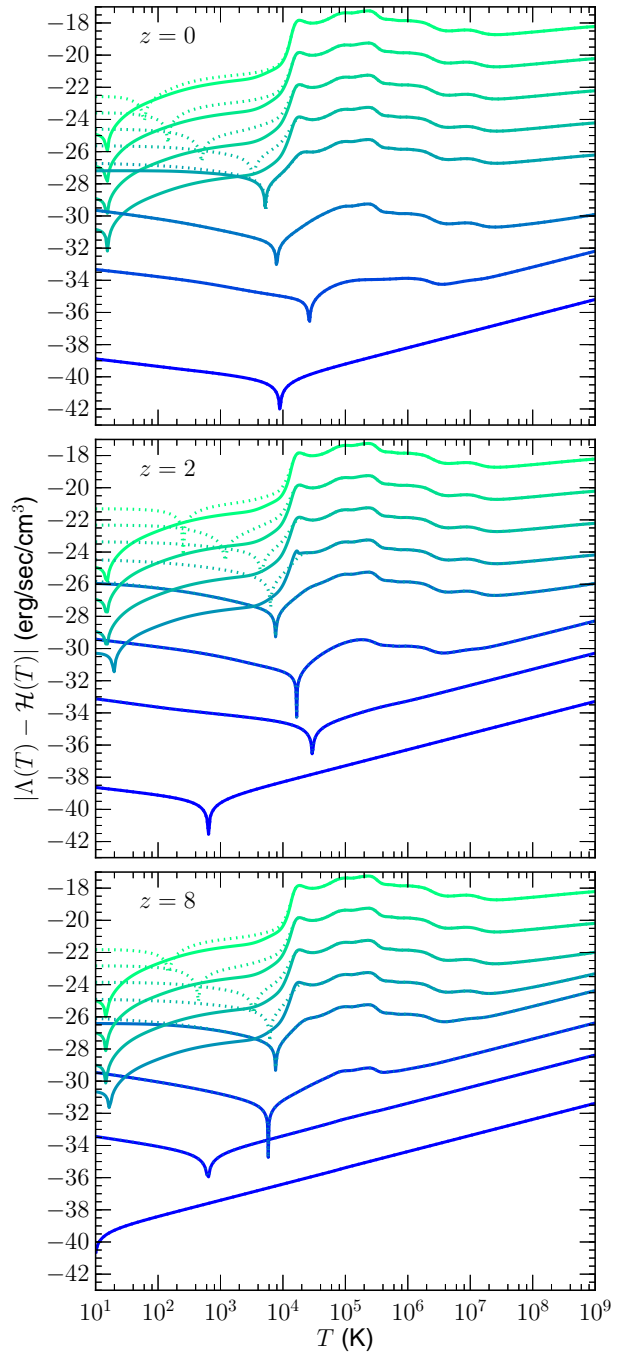}
\caption{Logarithm of the net cooling rate of solar metallicity gas
  exposed to the $z=0$ UVB (top panel), the $z=2$ UVB (middle panel),
  and the weak $z=8$ UVB (bottom panel) with self-shielding (full
  lines) and without self-shielding (dotted lines). The line colors
  reflect gas density (from the top curve downwards:~$n_{\sf H}=100$,
  $10$, $1$, $10^{-1}$, $10^{-2}$, $10^{-4}$, $10^{-6}$, and
  $10^{-9}$~cm$^{-3}$).
\label{fig:shielding}}
\end{figure}

In Fig. \ref{fig:shielding}, we plot the net cooling rates,
  defined as $\Lambda_{\sf net} = \left| \Lambda-{\cal H} \right|$, of
  solar metallicity gas exposed to the $z=0$ (top panel), $z=2$
  (middle panel), and $z=8$ (bottom panel) UVB for different
  densities, with self-shielding (full lines) and without
  self-shielding (dotted lines). Clearly, for densities below the
self-shielding threshold density of $n_{\rm HI, crit}=0.007$~cm$^{-3}$
switching shielding on or off makes no difference:~the UVB can fully
penetrate, ionize, and heat the gas. Above the self-shielding
threshold (roughly the top four curves in each panel), the gas is
dense enough to recombine and, with self-shielding switched on, to
strongly suppress the UVB. As a consequence, the heating rate drops
steeply below $T \sim 10^4$~K. With self-shielding switched off,
Hydrogen also largely recombines but now does not suppress the
UVB. This has a profound influence on the heating rate which keeps on
increasing towards lower temperatures until reaching a plateau below
$T \sim 10^4$~K, as discussed earlier. Without self-shielding against
the UV radiation, ionization levels tend to be higher, leading to
higher abundances of important cooling ions such as C{\sc ii} and,
consequently, to higher cooling rates in the $T < 10^4$~K temperature
regime.  Self-shielding also has a strong effect on the
  equilibrium temperature of the gas. If, without self-shielding, the
  UVB can flood the gas unimpeded, the equilibrium temperature can be
  over two orders of magnitude higher than in the self-shielded case
  (e.g. $T_{\sf eq}=6500$~K without self-shielding versus $T_{\sf
    eq}=20$~K with self-shielding for the $z=2$ UVB at gas density
  $n_{\rm HI}=0.1$~cm$^{-3}$. 

The most striking effect of the self-shielding prescription is the
equilibrium temperature to which the gas would evolve, given
sufficient time.  The sudden downward break in each curve
  $\Lambda_{\sf net}$ indicates where $\Lambda={\cal H}$ and hence
  marks the equilibrium temperature.  In the middle panel of
Fig. \ref{fig:shielding}, the strong $z=2$ UVB is employed while in
the top and bottom panels, the results for the much weaker $z=0$ and
$z=8$ UVBs are shown. Without self-shielding, gas is incapable of
radiatively cooling significantly below $T \sim 10^4$~K, except for
sufficiently high densities and at large enough redshifts where the
UVB is still weak. Except at late and at very early cosmic times, the
UVB severely inhibits the formation of the cold, high-density clouds
in which stars are thought to form while it facilitates the removal of
low-density gas by ram-pressure stripping and galaxy interactions
\citet{may07,go10}. Together with supernova feedback, which most
strongly affects high-density regions, this almost completely
extinguishes star formation in simulated dwarf galaxies after $z \sim
2$ \citep{si12}. However, while the specific star-formation rate of
Local Volume dwarf galaxies was generally larger before $z \sim 2$
compared with later times, they do show a wide variety of more or less
continuous star-formation histories over their full lifetimes
\citep{wi11}. It remains to be seen whether a proper, self-consistent
treatment of the effects of photo-heating by the UVB, including the
effects of self-shielding, on the ionization equilibrium and the
resultant cooling and heating rates, can alleviate this problem.

\begin{figure}
\centering
\includegraphics[width=0.5\textwidth]{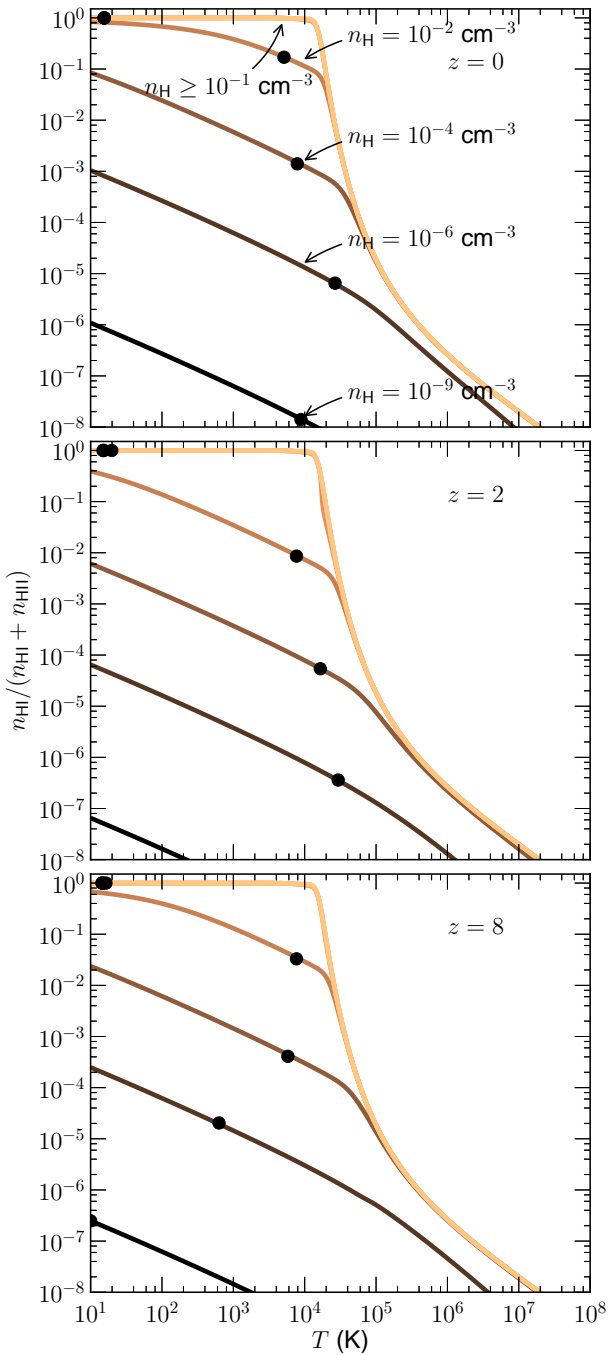}
\caption{Neutral fraction of solar metallicity gas exposed to the
  $z=0$ UVB (top panel), the $z=2$ UVB (middle panel), and the weak
  $z=8$ UVB (bottom panel), as a function of temperature and density
  (indicated in the figure, from the top curve downwards:~$n_{\sf
    H}=100$, $10$, $1$, $10^{-1}$, $10^{-2}$, $10^{-4}$,
  $10^{-6}$, and $10^{-9}$~cm$^{-3}$). The black dots indicate the
  neutral fraction at the equilibrium temperature for each density and
  redshift.
\label{fig:neutral}}
\end{figure}

Moreover, the non-inclusion of self-shielding makes it virtually
impossible to clearly identify different phases in the ISM of a
simulated galaxy and to compare them with observed galaxies. For
instance, in their analysis of a fully cosmological hydrodynamical
dwarf galaxy which was simulated including an UVB but neglecting
self-shielding, \citet{pi11} find that gas particles typically have
temperatures of the order of $7,000-9,000$~K. In order to investigate
the simulated dwarf's H{\sc i} properties, these authors are forced to
select as ``cold'' gas particles those with temperatures below
$15,000$~K. Since the calculation of the cooling and heating curves
presented here involves the determination of the ionization
equilibrium as a function of the gas properties, we immediately have
the neutral Hydrogen fraction at our disposal. As can be seen
Fig. \ref{fig:neutral}, in the presence of a UVB, the neutral fraction
of a gas parcel is not only temperature dependent but also strongly
density and redshift dependent. The black dots in this figure indicate
the neutral fraction at the equilibrium temperature for each redshift
and gas density. Clearly, the Hydrogen gas has to be denser than
several $0.01$~cm$^{-3}$ before it can become neutral (i.e. $n_{\rm
  HI}/n_{\rm H} \gtrsim 0.95$). This broadly agrees with the estimated
lower bound on the density of the Warm Neutral Medium in the Milky Way
\citep{wo03}.

\section{Discussion} \label{sect:conc}

In this paper, we have presented a new calculation of composition
dependent cooling and heating curves intended for use in numerical
simulations of galaxy formation and evolution. For each elemental mix,
density, temperature, and cosmological time the ionization equilibrium
was determined using a modified version of ChiantiPy, equipped with
collisional and radiative ionization, by the cosmic UV background and
an interstellar radiation field, and charge-transfer reactions. We
believe these curves address several drawbacks of currently available
tabulations of cooling rates.

We have shown that the full range of abundance variations likely to be
encountered in stars and neutral and ionized gas in a galaxy can be
described adequately by a simplified chemical-evolution model in which
there are only two contributions to the elemental yields:~a ``fast''
one (encompassing the contributions from SN{\sc ii} and massive IMS)
and a ``slow'' one (encompassing the contributions from SN{\sc i}a and
less massive IMS). The ratio of both contributions can be linked
directly to the Fe and Mg abundances which provide us with two strong
handles on {\em all} other element abundances. 

Thus, the cooling and heating curves depend on only five parameters
(temperature, density, redshift, [Fe/H], and [Mg/Fe]). They are easily
tabulated, and can be efficiently interpolated during a simulation. We
have implemented a five-dimensional interpolator in our own simulation
code enabling us to employ the cooling and heating curves presented
here in galaxy evolution simulations. A detailed analysis of the
effects of using these new curves on such simulations, especially
regarding the evolution of the star-formation rate in dwarf galaxies,
will be presented in a forthcoming paper.

We believe that the main advantage of this work is that, within the
confines of a well-defined chemical evolution model and adopting the
ionization equilibrium approximation, it provides accurate cooling and
heating curves for a wide range of physical and chemical gas
properties. These should be valid as long as the gas is neutral or,
partially, ionized (molecule and dust formation have not been taken
into account). Moreover, during a numerical simulation, one need only
follow the evolution of the Fe and Mg abundances, leading to a
reduction in memory requirements and computational cost compared to
the element-by-element approach. It takes into account self-shielding
in an approximate but motivated and well-tested way. This is key to
resolving the formation of cold, neutral, high-density clouds suitable
for star formation and to studying the structure of the multi-phase
ISM in galaxy simulations. Moreover, since we have stored the
ionization equilibrium for each combination of temperature, density,
redshift, [Fe/H], and [Mg/Fe], we can in principle calculate any
desired physical property of the gas using ChiantiPy.

Pre-compiled tables of the cooling and heating curves are available to
the community as online-only supporting information. These
  tables, future updates, and new material can also be found on and
  downloaded from http://users.ugent.be/$^{_\sim}$sdrijcke.

\section*{Acknowledgements}

We would like to thank the anonymous referee for his/her supportive
comments and suggestions. They improved the content and readability of
the paper.

\label{lastpage}


\begin{thebibliography}{}
\bibitem[Aubert \& Teyssier(2010)]{at10} Aubert D. \& Teyssier T.,
  2010, ApJ, 724, 244-266
\bibitem[Bekki(2013)]{b13} Bekki K., 2013, arXiv1304.1633B
\bibitem[Brook et al.(2012)]{br12} Brook C. B., Stinson G., Gibson B. K.,
  Ro\v{s}kar R., Wadsley J., Quinn T., 2012, MNRAS, 419, 771-779
\bibitem[Chabrier(2003)]{cha} Chabrier G., 2003, ApJ, 586, L133-L136
\bibitem[Cloet-Osselaer et al.(2012)]{co12} Cloet-Osselaer A., De
  Rijcke S., Schroyen J., Dury V., 2012arXiv1203.1863C
\bibitem[Dere et al.(2009)]{de09} Dere K. P., Landi E., Young P. R.,
  Del Zanna G., Landini M., Mason H. E., 2009, A\&A, 498, 915-929
\bibitem[Faucher-Gigu\`ere et al.(2009)]{fg09} Faucher-Gigu\`ere
  C.-A., Lidz A., Zaldarriaga M., Hernquist L., 2009, ApJ, 703,
  1416-1443
\bibitem[Faucher-Gigu\`ere et al.(2010)]{f10} Faucher-Gigu\`ere C.-A., Kere\v{s} D., Dijkstra M.,
  Hernquist L., Zaldarriaga M, 2010, ApJ, 725, 633-657
\bibitem[Fitzpatrick(2010)]{fi10} Fitzpatrick E. L., 2010, ApJ, 725, 2401-2416
\bibitem[Fran\c{c}ois et al.(2004)]{fr04} Fran\c{c}ois P., Matteucci
  F., Cayrel R., Spite M., Spite F., Chiappini C., 2004, A\&A, 421,
  613-621
\bibitem[Gabor \& Dav\'e(2012)]{gd12} Gabor J. M. \& Dav\'e R., 2012arXiv1202.5315G
\bibitem[Gavil\'an et al.(2005)]{ga05} Gavil\'an M., Buell J. F., Moll\'a M.,
2005, A\&A, 432, 861-877
\bibitem[Gnat \& Sternbert(2007)]{gs07} Gnat O. \& Sternberg A.,
  2007, ApJSS, 168, 213-230
\bibitem[Gnedin \& Hollon(2012)]{gh12} Gneding N. Y. \& Hollon N.,
  2012, ApJSS, 202, 13-20
\bibitem[Governato et al.(2010)]{go10} Governato F. Brook C., Mayer
  L., Brooks A., Rhee G., Wadsley J., Jonsson P., Willman B., Stinson
  G., Quinn T., Madau P., 2010, Nature, 463, 203-206
\bibitem[Grevesse et al.(2010)]{gr10} Grevesse N., Asplund M., Sauval
  A. J., Scott P., 2010, ApSS, 328, 179-183
\bibitem[Kim et al.(2012)]{ki12} Kim J.-H., Krumholz M. R., Wise
  J. H., Turk M. J., Goldbaum N. J., Abel T., 2012, arXiv1210.3361
\bibitem[Kingdon \& Ferland(1996)]{kf96} Kingdon J. B. \& Ferland
  G. J., 1996, ApJSS, 106, 205-211
\bibitem[Krumholz(2012)]{kr12} Krumholz M. R., 2012, ApJ, 759, 9-17
\bibitem[Landstreet(2011)]{la11} Landstreed J. D., 2011, A\&A, 528,
  A132
\bibitem[Maio et al.(2007)]{ma07} Maio U., Dolag K., Ciardi B.,
  Tornatore L., 2007, MNRAS, 379, 963-973
\bibitem[Mathis et al.(1983)]{ma83} Mathis J. S., Mezger P. G.,
  Panagia N., 1983, A\&A, 128, 212-229
\bibitem[Mayer et al.(2007)]{may07} Mayer L., Kazantzidis S.,
  Mastropietro C., Wadsley J., 2007, Nature, 445, 738-740
\bibitem[Nagamine et al.(2010)]{na10} Nagamine K., Choi J., Yajima H.,
  2010, ApJ, 725, L219-L222
\bibitem[Nomoto et al.(1997)]{no97} Nomoto K., Hashimoto M.,
  Tsujimoto T., Thielemann F.-K., Kishimoto N., Kubo Y., Nakasato
  N., 1997, NuPha, 616, 79-90
\bibitem[Pilkington et al.(2011)]{pi11} Pilkington K., Gibson B. K.,
  Calura F., Brooks A. M., Mayer L., Brook C. B., Stinson G. S.,
  Thacker R. J., Few C. G., Cunnama D., Wadsley J., 2011, MNRAS, 417,
  2891-2898
\bibitem[Revaz et al.(2009)]{re09} Revaz Y., Jablonka P., Sawala T.,
  Hill V., Letarte B., Irwin M., Battaglia G., Helmi A.,
  Shetrone M. D., Tolstoy E., Venn K. A., 2009, A\&A, 501, 189-206
\bibitem[Sawala et al.(2011)]{sa11} Sawala T., Guo Q., Scannapieco C.,
  Jenkins A., White S., 2011, MNRAS, 413, 659-668
\bibitem[Scannapieco et al.(2011)]{sc11} Scannapieco C., White S. D. M., Springel V., Tissera
  P. B., 2011, MNRAS, 417, 154-171
\bibitem[Schaye et al.(2010)]{sch10} Schaye J., Dalla Vecchia C.,
  Booth C. M., Wiersma R. P. C., Theuns T., Haas M. R., Bertone S.,
  Duffy A. R., McCarthy I. G., van de Voort F., 2010, MNRAS, 402,
  1536-1560
\bibitem[Schroyen et al.(2011)]{sch11} Schroyen J., De Rijcke S., Valcke
  S., Cloet-Osselaer A., Dejonghe H., 2011, MNRAS, 416, 601-617
\bibitem[Shen et al.(2010)]{sh10} Shen S., Wadsley J., Stinson G.,
  2010, MNRAS, 407, 1581-1596
\bibitem[Simpson et al.(2012)]{si12} Simpson C. M., Bryan G. L.,
  Johnston K. V., Smith B. D., Mac Low M.-M., Sharma S., Tumlinson J.,
  2012, arXiv:1211.1071v1
\bibitem[Smith, Sigurdsson, Abel(2008)]{ssa08} Smith B., Sigurdsson S., Abel
  T., 2008, MNRAS, 385, 1443-1454
\bibitem[Smith et al.(2011)]{sm11} Smith B. D., Hallman E. J., Shull
  J. M., O'Shea B. W., 2011, ApJ, 731, 6-26
\bibitem[Stancil et al.(1998)]{st98} Stancil P. C., Havener P. S., Krsti\'c
  P. S., Schultz D. R., Kimura M., Gu J.-P., Hirsch G., Buenker R. J.,
  Zygelman B., 1998, ApJ, 502, 1006-1009
\bibitem[Stancil et al.(1999)]{st99} Stancil P. C., Schultz D. R.,
  Kimura M., Gu J.-P., Hirsch G., Buenker R. J., 1999, A\&ASS, 140,
  225-234
\bibitem[Sutherland \& Dopita(1993)]{sd93} Sutherland R. S. \& Dopita
  M. A., 1993, ApJS, 88, 253-327
\bibitem[Tajiri \& Umemura(1998)]{tu98} Tajiri Y. \& Umemura M., 1998,
  ApJ, 502, 59-62
\bibitem[Tepper-Garc\'{\i}a et al.(2011)]{tp11} Tepper-Garc\'{\i}a T.,
  Richter P., Schaye J., Booth C. M., Dalla Vecchia C., Theuns T.,
  Wiersma R. P. C., 2011, MNRAS, 413, 190-212
\bibitem[Tolstoy, Hill, Tosi(2009)]{tht09} Tolstoy E., Hill V., Tosi
  M., 2009, ARA\&A, 47, 371-425
\bibitem[Tsujimoto et al.(1995)]{ts95} Tsujimoto T., Nomoto K., Yoshii
  Y., Hashimoto M., Yanagida S., Thielemann F.-K., 1995, MNRAS, 277,
  945-958
\bibitem[Vasiliev(2013)]{va13} Vasiliev E. O., 2013, MNRAS, 431,
  638-647
\bibitem[Verner et al.(1996)]{ve96} Verner D. A., Ferland G. J.,
  Korista K. T., Yakovlev D. G., 1996, ApJ, 465, 487-498
\bibitem[Weisz et al.(2011)]{wi11} Weisz D. R., Dalcanton J. J., Williams
  B. F., Gilbert K. M., Skillman E. D., Seth A. C., Dolphin A. E.,
  McQuinn K. B. W., Gogarten S. M., Holtzman J., Rosema K., Cole A.,
  Karachentsev I. D., Zaritsky D., 2011, ApJ, 739, article id. 5, 27
  pp.
\bibitem[Wiersma et al.(2009)]{wi09} Wiersma R. P. C., Schaye J.,
  Smith B. D., 2009, MNRAS, 393, 99-107
\bibitem[Wolfire et al.(2003)]{wo03} Wolfire M. G., McKee C. F.,
  Hollenbach D., Tielens A. G. G. M., 2003, ApJ, 587, 278-311
\bibitem[Worley et al.(2009)]{wo09} Worley C. C., Cottrell P. L.,
  Freeman K. C., Wylie-de Boer E. C., MNRAS, 400, 1039-1048
\bibitem[Worthey, Faber, Gonzalez(1992)]{wfg92} Worthey g., Faber
  S. M., Gonzalez J. J., 1992, ApJ, 398, 69-73
\bibitem[Yajima et al.(2011)]{ya11} Yajima H., Jun-Hwan C., Nagamine
  K., 2011, MNRAS, 412, 411-422
\end{thebibliography}
\end{document}